\documentclass[aps,preprintnumbers,superscriptaddress,showpacs]{revtex4}
\usepackage{epsfig}
\usepackage{psfrag}
\usepackage{amsfonts}
\usepackage{graphicx}
\usepackage{dcolumn}
\usepackage{bm}

\begin{document}

\title{Imaginary potential in strongly coupled $\mathcal{N}=4$ SYM plasma in a magnetic field}

\author{Zi-qiang Zhang}
\email{zhangzq@cug.edu.cn} \affiliation{School of mathematics and
physics, China University of Geosciences(Wuhan), Wuhan 430074,
China}

\author{De-fu Hou}
\email{houdf@mail.ccnu.edu.cn} \affiliation{Key Laboratory of
Quark and Lepton Physics (MOE), Central China Normal University,
Wuhan 430079,China}

\begin{abstract}
We study the effect of a constant magnetic field on the imaginary
part of a quarkonia potential in a strongly-coupled
$\mathcal{N}=4$ SYM plasma. We consider the pair axis to be
aligned perpendicularly and parallel to the magnetic field,
respectively. For both cases, we find that the presence of the
magnetic field tends to enhance the imaginary potential thus
decreasing the the thermal width. In addition, the magnetic field
has a stronger effect on the imaginary potential when the pair
axis is perpendicular to the magnetic field rather than parallel.

\end{abstract}

\maketitle
\section{Introduction}
The heavy ion collisions at RHIC and LHC have produced a new state
of matter so-called quark gluon plasma (QGP) \cite{JA,KA,EV}. One
experimental signal for QGP formation is dissociation of quarkonia
. It was suggested earlier that the main mechanism responsible for
this suppression is color screening \cite{TMA}. But recently some
authors argued that the imaginary part of the potential,
ImV$_{QQ}$, may be a more important reason than screening
\cite{ML,AB,NB,MAE}. Subsequently, this quantity has been studied
in weakly coupled theories, see e.g. \cite{NB1,AD,MM,VC}. However,
much experiment data indicates that QGP is strongly coupled
\cite{EV}, so it would be interesting to study the imaginary
potential in strongly coupled theories with the aid of
nonpetubative methods. Such methods are now available via the
AdS/CFT correspondence
\cite{Maldacena:1997re,Gubser:1998bc,MadalcenaReview}.

AdS/CFT, the duality between a string theory in AdS space and a
conformal field theory in the physical space-time, has yielded
many important insights for studying different aspects of QGP
\cite{JCA}. In this approach, Noronha and Dumitru have studied the
imaginary potential of quarkonia for $\mathcal{N}=4$ SYM theory in
their seminal work \cite{JN}. Therein, the ImV$_{QQ}$ is related
to the effect of thermal fluctuations due to the interactions
between the heavy quarks and the medium. After \cite{JN}, there
were many attempts to address ImV$_{QQ}$ in this direction, for
instance, the ImV$_{QQ}$ of static quarkonia is studied in
\cite{JN1,KB}. The effect of velocity on ImV$_{QQ}$ is discussed
in \cite{JN2,MAL}. The finite 't Hooft coupling corrections on
ImV$_{QQ}$ is analyzed in \cite{KB1}. The influence of chemical
potential on ImV$_{QQ}$ is investigated in \cite{ZQ}. For study of
ImV$_{QQ}$ in some AdS/QCD models, see \cite{NR,JS}. Moreover,
there are other ways to study ImV$_{QQ}$ from AdS/CFT, see
\cite{JLA,THA}.

Recently, there are various observables or quantities that have
been studied in strongly coupled $\mathcal{N}=4$ SYM plasma under
the influence of a magnetic field, such as entropy density
\cite{ED}, conductivity \cite{KAMA}, shear viscosity to entropy
density ratio \cite{RCR}, heavy quark potential \cite{RRO}, drag
force \cite{KIM} and jet quenching parameter \cite{SLI}. Motivated
by this, in this paper we study the effect of a constant magnetic
field on the imaginary part of heavy quarkonia potential in a
strongly-coupled $\mathcal{N}=4$ SYM plasma. Specifically, we
would like to see how a constant magnetic field affects ImV$_{QQ}$
in this case. This is the purpose of the present work.

The rest of the paper is as follows. In the next section, we
briefly review the background metric in the presence of a magnetic
field given in \cite{ED}. In section 3, we introduce the numerical
procedure and show some numerical solutions. In section 4, we
study the effect of a magnetic field on the imaginary potential
for the pair axis to be aligned perpendicularly and parallel to
the magnetic field, in turn. The last part is devoted to
conclusion and discussion.


\section{magnetic brane background}
Let us begin with a briefly review of the holographic dual of
$\mathcal{N}=4$ SYM theory in the presence of a magnetic field
\cite{ED}. The holographic model is Einstein gravity coupled with
a Maxwell field, corresponding to strongly coupled $\mathcal{N}=4$
SYM subjected to a constant and homogenous magnetic field. The
bulk action is given by

\begin{equation}
S=\frac{1}{16\pi G_5}\int
d^5x\sqrt{-g}(R+\frac{12}{L^2}-F^2)+S_{body},\label{action}
\end{equation}
where $G_5$ denotes the 5-dimensional gravitational constant, $L$
stands for the radius of the asymptotic $AdS_5$ spacetime. $F$
refers to the Maxwell field strength 2-form. The boundary term
$S_{body}$ contains the Chern-Simons terms, Gibbons-Hawking terms
and other contributions necessary for a well posed variational
principle, but all of them do not affect the solutions considered
here \cite{ED}.

The equations of motion obtained from (\ref{action}) are
\begin{equation}
R_{\mu\nu}+\frac{4}{L^2}g_{\mu\nu}+\frac{1}{3}F_{\rho\sigma}F^{\rho\sigma}g_{\mu\nu}-2F_{\mu\rho}F_\nu^\rho=0,
\label{ein}
\end{equation}
and the Maxwell's field equations
\begin{equation}
\nabla_\mu F^{\mu\nu}=0.
\end{equation}

For simplicity, we set $L=1$ from now on. Then a general ansatz
for the magnetic brane geometry can be written as
\begin{equation}
ds^2=-H(r)dt^2+e^{2P(r)}(dx^2+dy^2)+e^{2K(r)}dz^2+\frac{dr^2}{H(r)},\label{metric}
\end{equation}
with
\begin{equation}
F=Bdx\wedge dy,
\end{equation}
where the boundary is located at $r=\infty$ and the horizon is
located at $r=r_h$ with $H(r_h)=0$. The constant $B$ stands for
the bulk magnetic field, oriented along the $z$ direction. Also,
$H(r)$, $P(r)$ and $K(r)$ can be determined by solving the
equations of motion.

From (\ref{metric}), the equations of motion read
\begin{equation}
H(P^{\prime\prime}-K^{\prime\prime})+(H^\prime+H(2P^\prime+K^\prime))(P^\prime-K^\prime)=-2B^2e^{-4P}\label{e1},
\end{equation}
\begin{equation}
2P^{\prime\prime}+K^{\prime\prime}+2(P^\prime)^2+(K^\prime)^2=0,\label{e2}
\end{equation}
\begin{equation}
\frac{1}{2}H^{\prime\prime}+\frac{1}{2}H^{\prime}(2P^\prime+K^\prime)=4+\frac{2}{3}B^2e^{-4P},\label{e3}
\end{equation}
\begin{equation}
2H^\prime P^\prime+H^\prime K^\prime+2H(P^\prime)^2+4HP^\prime
K^\prime=12-2B^2e^{-4P},\label{e4}
\end{equation}
where $H\equiv H(r)$, $P\equiv P(r)$, $K\equiv K(r)$ and the
derivations are with respect to $r$. For the above equations, one
can check that with $B=0$, the solution reduces to that of
$AdS_5$, represented by $H=e^{2P}=e^{2K}=r^2$. While for non-zero
$B$, one finds an exact solution in the asymptotic IR regime,
represents the product of a BTZ black hole times a two dimensional
tours $T^2$ in the spatial directions orthogonal to the magnetic
field (the $x,y$ directions are compact), as
\begin{equation}
ds^2=-3(r^2-r_h^2)dt^2+\frac{B}{\sqrt{3}}(dx^2+dy^2)+3r^2dz^2+\frac{dr^2}{3(r^2-r_h^2)}.
\label{metric1}
\end{equation}

It should be noticed that the metric (\ref{metric1}) is valid only
near the horizon ($r\sim r_h$), where the scale is much smaller
than the magnetic field, $r\ll\sqrt{B}/3$, and recently some
authors have used it to study the effect of a strong magnetic
field on the drag force \cite{KIM} and jet quenching parameter
\cite{SLI}.

However, when discuss the effect of a general magnetic field, not
restricted to a strong magnetic field (or in the IR regime), one
should use a solution that interpolates between (\ref{metric1}) at
small $r$ and $AdS_5$ at large $r$. As discussed in \cite{ED},
this represents an RG flow between a $D=3+1$ CFT in the infrared
and a $D=1+1$ CFT in the ultraviolet. Unfortunately, no analytic
solution can be found in this case. Therefore, one needs to turn
to numerical methods. In the next section, we will follow the
argument in \cite{ED} to present the numerical procedure.

\section{Numerical solutions}
To begin with, we derive some useful equations. After eliminating
the $B^2e^{-4P}$ terms in (\ref{e1})$-$(\ref{e4}), we have
\begin{equation}
3H^{\prime\prime}+5(K^\prime+2P^\prime)H^\prime+4({P^\prime}^2+2P^\prime
K^\prime)H-48=0,\label{e5}
\end{equation}
\begin{equation}
3HP^{\prime\prime}+2H{P^\prime}^2-H^\prime P^\prime-5HP^\prime
K^\prime+12-2H^\prime K^\prime=0,\label{e6}
\end{equation}
\begin{equation}
3HK^{\prime\prime}+3H{K^\prime}^2+4H^\prime K^\prime+10HP^\prime
K^\prime+2H{P^\prime}^2+2H^\prime P^\prime-24=0,\label{e7}
\end{equation}
before solving the above equations, it is convenient to rescale
the radial coordinates as follows \cite{ED}: First, we rescale $t$
and $r$ as $r \rightarrow\bar{r}$ and $t\rightarrow\bar{t}$, and
fix the horizon at $\bar{r}_h=1$, so that
\begin{equation}
H(1)=0,\qquad H^{\prime}(1)=1. \label{h}
\end{equation}

In this case, the Hawking temperature is given by
\begin{equation}
T=\frac{\sqrt{-g^\prime_{\bar{t}\bar{t}}g^{\bar{r}\bar{r}\prime}}}{4\pi}\big{|}
_{\bar{r}=1}=\frac{1}{4\pi}.
\end{equation}

Also, we rescale $x$, $y$, $z$ coordinates as
\begin{equation}
P(1)=K(1)=0,\qquad P^\prime(1)=4-\frac{b^2}{3},\qquad
K^\prime(1)=4+\frac{b^2}{6}, \label{w}
\end{equation}
where $b$ refers to the value of the magnetic field in the
rescaled coordinates. Note that the second equation in (\ref{w})
implies that, if one takes $b>2\sqrt{3}$, the term $P^\prime(1)$
will be negative, indicating that the geometry will not be
asymptotically $AdS_5$. Therefore, we take $0\leq b<2\sqrt{3}$
here.

On the other hand, the geometry has the asymptotic behavior as
$\bar{r}\rightarrow \infty$, so that
\begin{equation}
H(\bar{r})\rightarrow \bar{r}^2, \qquad e^{2P(\bar{r})}\rightarrow
m(b) \bar{r}^2, \qquad e^{2K(\bar{r})}\rightarrow p(b) \bar{r}^2,
\end{equation}
where $m(b)$ and $p(b)$ are rescaling parameters which can be
obtained numerically. In addition, the physical magnetic field
$\mathcal{B}$ can be written as
\begin{equation}
\mathcal{B}=\sqrt{3}\frac{b}{m(b)}.\label{b}
\end{equation}

In fact, one can also derive the interval of $b$ from
Eq.(\ref{b}). One can numerically check that $m(b)$ is a
decreasing function of $b$ and $m(b\rightarrow
2\sqrt{3})\rightarrow 0$ (see the left panel of fig.1). Thus,
$b\in[0,2\sqrt{3})$ leads to $\mathcal{B}\in[0,\infty)$. Namely,
one can cover in practice all values of $\mathcal{B}$ for $0\leq
b<2\sqrt{3}$. Here it should be noticed that the actual
$\mathcal{B}$ is immaterial, due to the presence of any other
scale in the theory besides it. However, when switching on the
temperature, this situation changes. In that case, one finds a new
dimensionless scale given by $\mathcal{B}/T^2$ \cite{RRO}.
Therefore, one can tune the anisotropy in the imaginary potential
by varying the value of the magnetic field.

Now we are ready to numerically solve the equations
(\ref{e5})$-$(\ref{e7}). For simplicity, we delete from now on the
bars in the rescaled coordinates. The numerical procedures are as
follows:

1). Choosing a value of $b$, one solves the coupled equations
(\ref{e5})$-$(\ref{e7}) with the conditions (\ref{h}) and
(\ref{w}).

2). Fitting the asymptotic data for $e^{2P(r)}\rightarrow m(b)
r^2$ and $e^{2K(r)}\rightarrow p(b) r^2$, one gets the values of
$m(b)$ and $p(b)$. Meanwhile, the value of $\mathcal{B}$ is
determined.

3). To go back to the original coordinates, one sets
$(e^{2P(r)},e^{2K(r)})\rightarrow
(e^{2P(r)}/m(b),e^{2K(r)}/p(b))$, after this, the numerical
solutions can be obtained.

4). Likewise, one can discuss other cases by varying the value of
$b$.

Here we present some numerical solutions. In the left panel of
fig.1, we plot $m(b)$, $p(b)$ versus $b$, and we have checked that
it matches the fig.1 in \cite{RCR}. Also, in the right panel of
fig.1, we plot $ln H(r)$, $K(r)$, $P(r)$ versus $r$ for $b=2.7$,
and we find that it is completely consistent with the fig.3 in
\cite{RRO}.
\begin{figure}
\centering
\includegraphics[width=8cm]{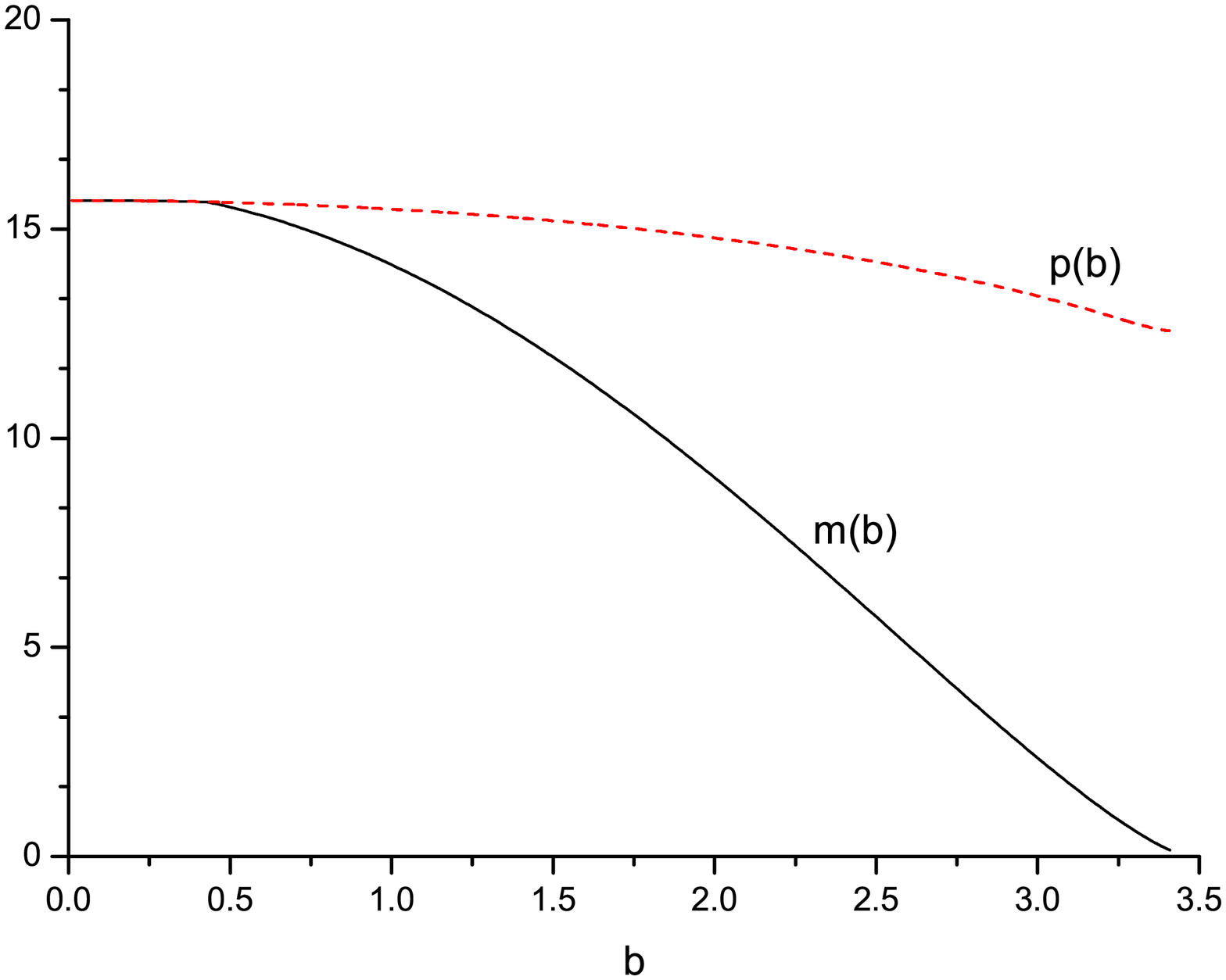}
\includegraphics[width=8cm]{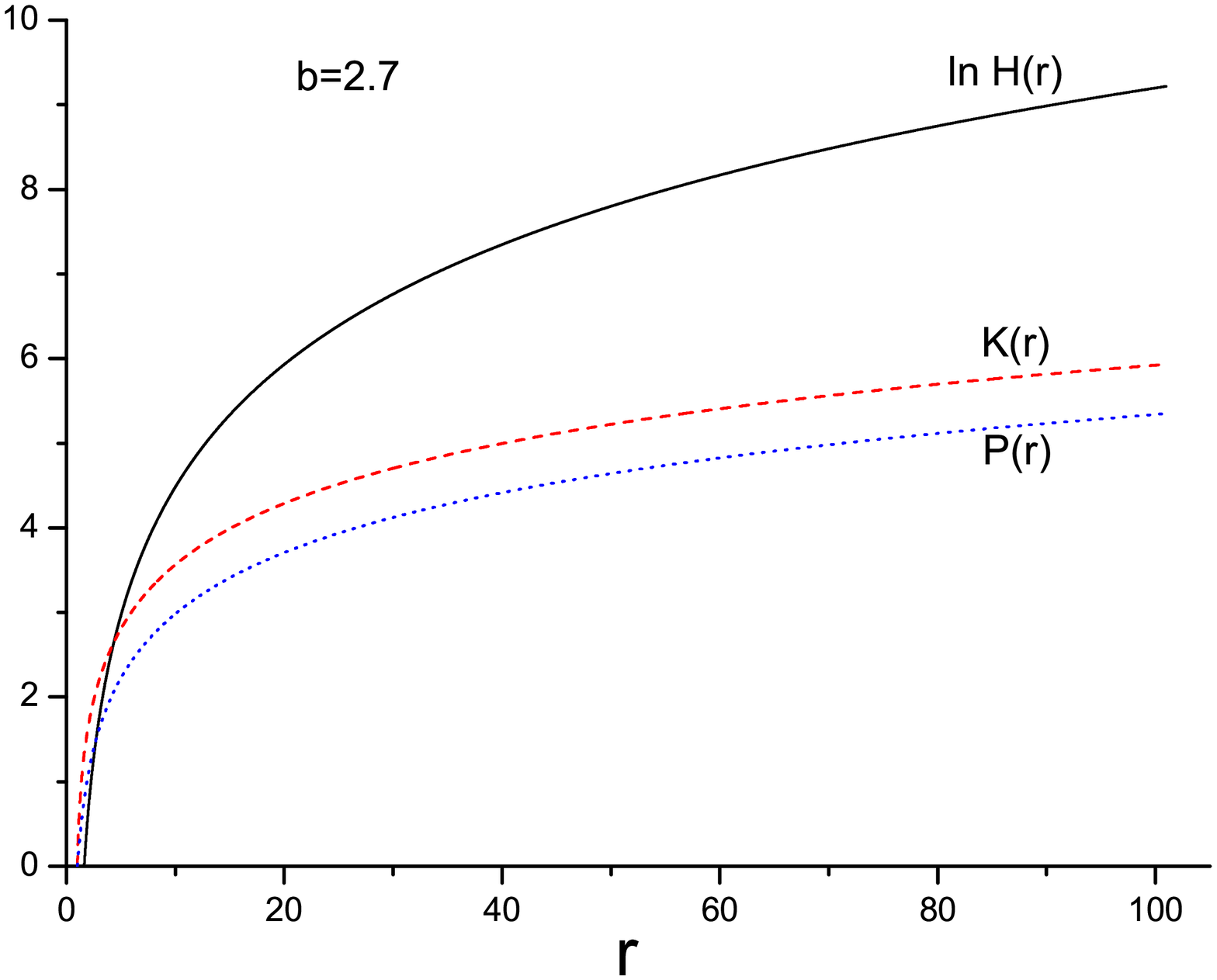}
\caption{Left: $p(b)$ (dash curve) and $m(b)$ (solid curve) versus
$b$. Right: $ln H(r)$ (solid curve), $K(r)$ (dash curve) and
$P(r)$ (dot curve) versus $r$ for $b=2.7$.}
\end{figure}

\section{imaginary potential}
Now we investigate the ImV$_{QQ}$ of a heavy quarkonia for the
background metric (\ref{metric}). Generally, to analyze the effect
of a magnetic field, one needs to consider different orientations
for the pair axis with respect to the magnetic field, i.e.,
transverse ($\theta=\pi/2$), parallel ($\theta=0$), and arbitrary
direction ($\theta$). In this work, we study two extreme cases:
$\theta=\pi/2$ and $\theta=0$.

\subsection{Transverse to the magnetic field ($\theta=\pi/2$)}
Considering the $Q\bar{Q}$ axis perpendicularly to the magnetic
field in the $x$ direction, the coordinates are parameterized as
\begin{equation}
t=\tau, \qquad x=\sigma,\qquad y=0,\qquad z=0,\qquad
r=r(\sigma),\label{par}
\end{equation}
where the quark and anti-quark are located at $x=-\frac{D}{2}$ and
$x=\frac{D}{2}$, respectively. Here $D$ is the inter-distance of
$Q\bar{Q}$. The configuration of the string world-sheet is
presented in fig.2.

To proceed, the string action is given by
\begin{equation}
S=-\frac{1}{2\pi\alpha^\prime}\int d\tau d\sigma\mathcal
L=-\frac{1}{2\pi\alpha^\prime}\int d\tau d\sigma\sqrt{-g},
\label{S}
\end{equation}
with $g$ the determinant of the induced metric and
\begin{equation}
g_{\alpha\beta}=g_{\mu\nu}\frac{\partial
X^\mu}{\partial\sigma^\alpha} \frac{\partial
X^\nu}{\partial\sigma^\beta},
\end{equation}
where $g_{\mu\nu}$ and $X^\mu$ represent the metric and target
space coordinates, respectively. The parameter $\alpha^\prime$ is
related to 't Hooft coupling as $1/\alpha^\prime=\sqrt{\lambda}$.

Substituting (\ref{par}) into (\ref{metric}), the induced metric
becomes
\begin{equation} g_{00}=H(r), \qquad
g_{01}=g_{10}=0,\qquad
g_{11}=e^{2P(r)}+\frac{1}{H(r)}(\frac{dr}{d\sigma})^2.
\end{equation}

Then one can identify the lagrangian density as
\begin{equation}
\mathcal L=\sqrt{H(r)e^{2P(r)}+(\frac{dr}{d\sigma})^2},
\end{equation}

Notice that the action does not depend on $\sigma$ explicitly, so
$\mathcal L$ satisfies
\begin{equation}
\mathcal L-\frac{\partial\mathcal L}{\partial
(\frac{dr}{d\sigma})}\frac{dr}{d\sigma}=\frac{H(r)e^{2P(r)}}{\sqrt{H(r)e^{2P(r)}+(\frac{dr}{d\sigma})^2}}=constant.
\end{equation}

Imposing the boundary condition at $\sigma=0$ (the deepest point
of the U-shaped string),
\begin{equation}
\frac{dr}{d\sigma}=0,\qquad r=r_*,
\end{equation}
one finds
\begin{equation}
\frac{dr}{d\sigma}=\sqrt{\frac{a^2(r)-a(r)a(r_*)}{a(r_*)}}\label{dotr},
\end{equation}
with
\begin{equation}
a(r)=H(r)e^{2P(r)},\qquad a(r_*)=H(r_*)e^{2P(r_*)}.
\end{equation}

Integrating (\ref{dotr}), the inter-distance of $Q\bar{Q}$ reads
\begin{equation}
D=2\int_{r_*}^{\infty}dr\frac{d\sigma}{dr}=2\int_{r_*}^{\infty}dr\sqrt{\frac{a(r_*)}{a^2(r)-a(r)a(r_*)}}\label{x}.
\end{equation}
\begin{figure}
\centering
\includegraphics[width=5cm]{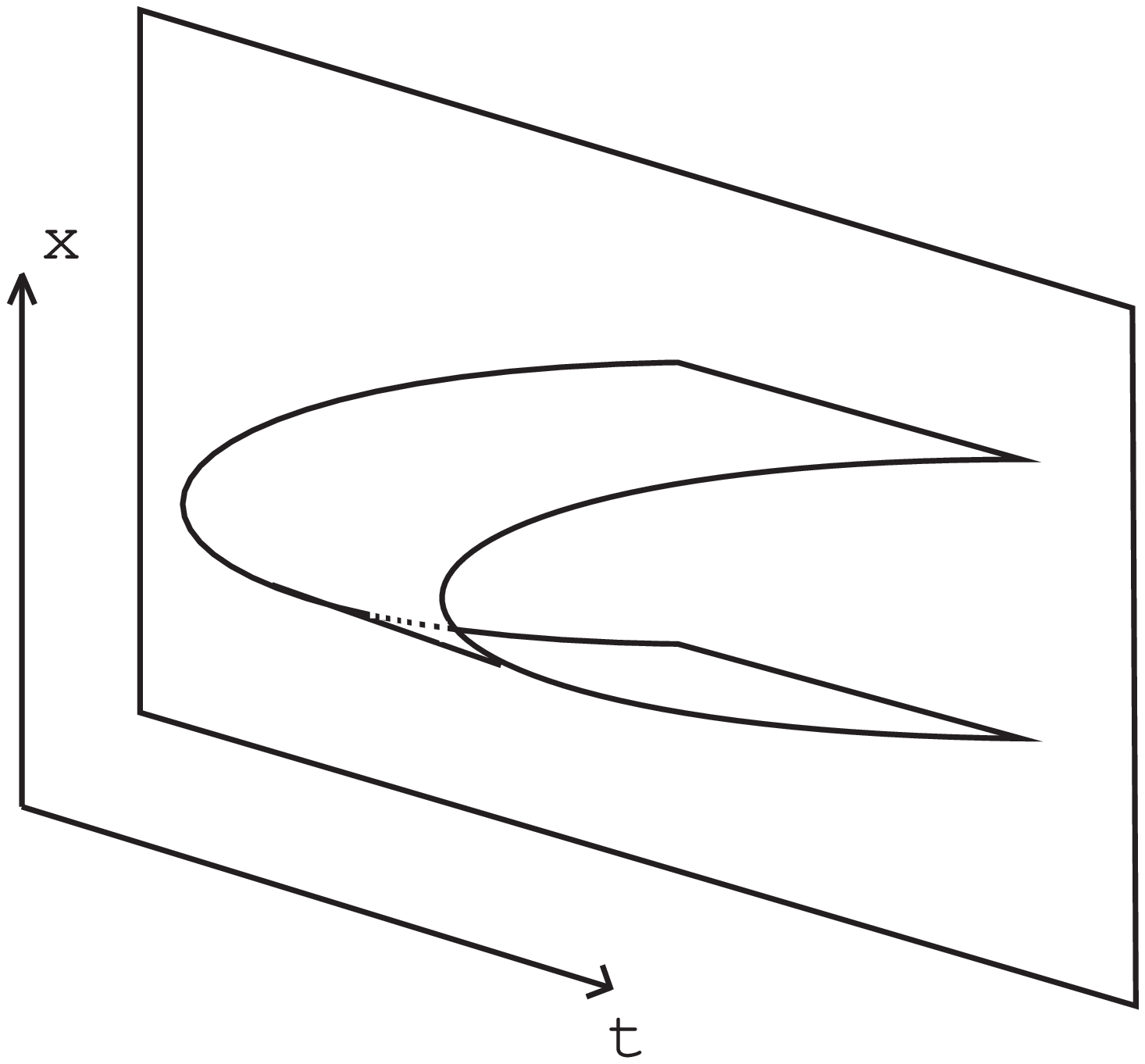}
\includegraphics[width=8cm]{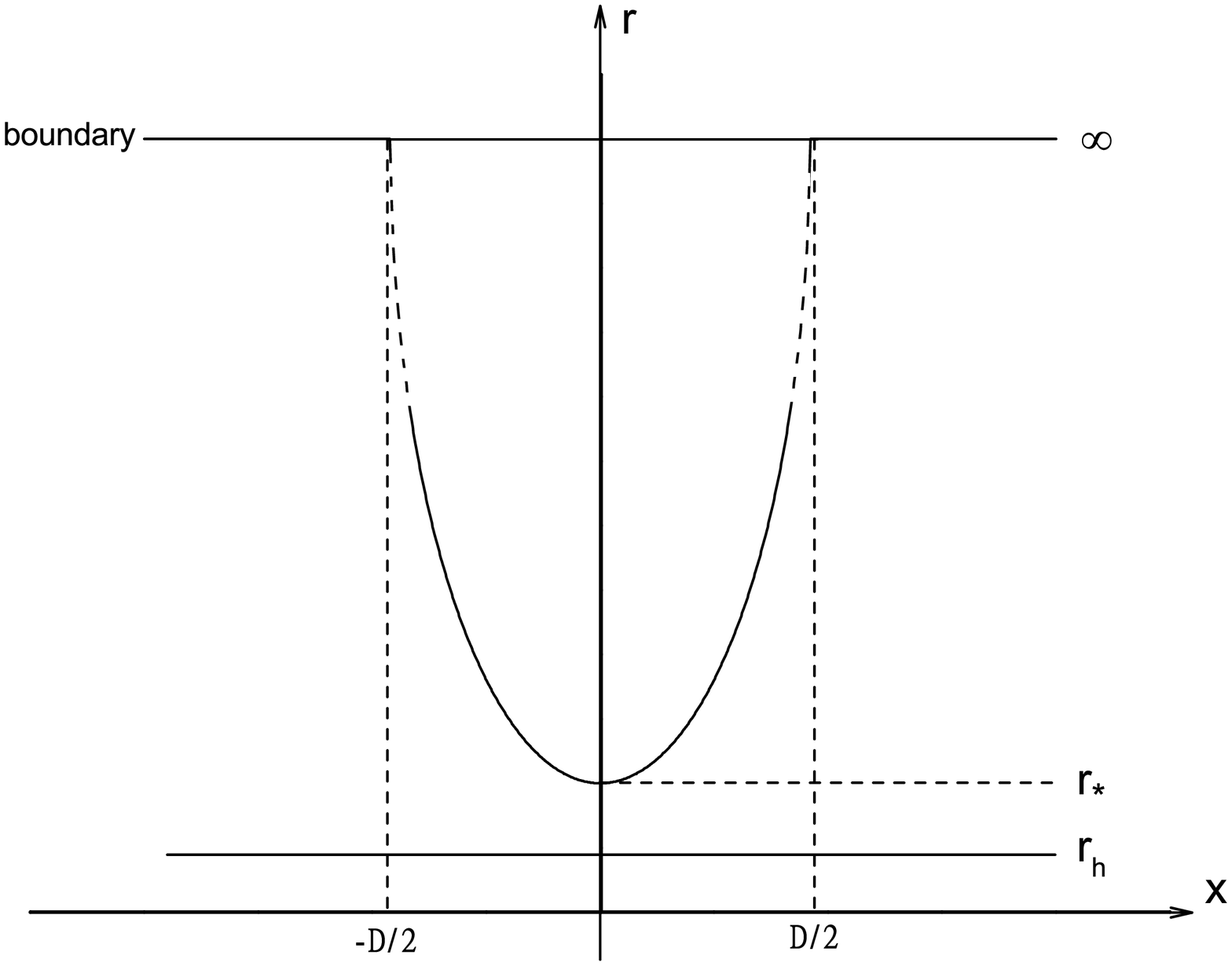}
\caption{The configuration of the string world-sheet.}
\end{figure}

Substituting (\ref{dotr}) into (\ref{S}), the action for the quark
pair is obtained as
\begin{equation}
S=\frac{\mathcal{T}}{\pi\alpha^\prime}\int_{r_h}^{\infty}dr\sqrt{\frac{a(r)}{a(r)-a(r_*)}},
\end{equation}
note that this action contains the self-energy contributions from
the free $Q\bar{Q}$ pair which, themselves, are divergent. To
obtain the $Q\bar{Q}$ interaction potential (or the quark
potential), one needs to cure this divergence by subtracting from
$S$ the action $S_0$ of a free $Q\bar{Q}$ pair \cite{JMM,ABR,SJR},
given by
\begin{equation}
S_0=\frac{\mathcal{T}}{\pi\alpha^\prime}\int_{r_h}^{\infty}dr\sqrt{g_{tt}g_{rr}}=\frac{\mathcal{T}}{\pi\alpha^\prime}\int_{r_h}^{\infty}dr.
\end{equation}

As a result, the heavy quark potential is found to be
\begin{equation}
V_{Q\bar{Q}}=\frac{S-S_0}{\mathcal{T}}=\frac{1}{\pi\alpha^\prime}\int_{r_*}^{\infty}dr[\sqrt{\frac{a(r)}{a(r)-a(r_*)}}-1]-
\frac{1}{\pi\alpha^\prime}\int_{r_h}^{r_*}dr\label{re},
\end{equation}
actually, this quantity has been studied for the background metric
(\ref{metric}) in \cite{RRO}, and the results show that the
presence of a magnetic field increases the heavy quark potential
or weakens the attraction between the heavy quarks.

To proceed, we study the imaginary potential by using the thermal
worldsheet fluctuation method \cite{JN,JN1}. To extract
ImV$_{QQ}$, one considers the effect of thermal world sheet
fluctuations $\delta r(x)$ around the classical configurations
$r_c(x)$,
\begin{equation}
r(x)=r_c(x)\rightarrow r(x)=r_c(x)+\delta r(x),
\end{equation}
where the boundary condition is $\delta r(\pm D/2)=0$. Note that
$r_c(x)$ solves $\delta S_{NG}=0$. For the sake of simplicity,
$\delta r(x)$ is taken to be of arbitrarily long wavelength, i.e.,
$\frac{d\delta r(x)}{dx}\rightarrow0$. In other words, the
fluctuations at each string point are independent functions if one
considers the long wavelength limit. The physical picture of the
thermal fluctuations is shown in fig.3.

\begin{figure}
\centering
\includegraphics[width=4cm]{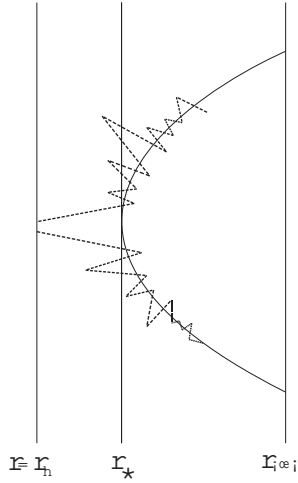}
\caption{The effect of thermal fluctuations (dashed line) around
the classical configuration (solid line). If $r_*$ is close enough
to $r_h$, the fluctuations of very long wavelength may reach the
horizon.}
\end{figure}

Then the string partition function that takes into account the
fluctuations can be written as
\begin{equation}
Z_{str} \sim \int \mathcal{D}\delta r(x)e^{iS_{NG}(r_c(x)+\delta
r(x))}.
\end{equation}

By dividing the interval $-D/2<X<D/2$ into $2N$ points
$x_j=j\Delta(x)$ with $j=-N,-N+1,...,N$ and $\Delta x\equiv
D/(2N)$, one finds
\begin{equation}
Z_{str} \sim \lim_{N\to\infty} \int d[\delta r(x_{-N})]\cdots
d[\delta r(x_{N})]exp[\frac{i\mathcal{T}\Delta
x}{2\pi\alpha^\prime}\sum_{j}\sqrt{(r^\prime_j)^2+a(r_j)}],\label{zs}
\end{equation}
where $r_j\equiv r(x_j)$ and $r_j^\prime\equiv r^\prime(x_j)$. The
thermal fluctuations are more important around $x=0$, where
$r=r_*$. Therefore, it is reasonable to expand $r_c(x_j)$ around
$x=0$, keeping only terms up to second order in $x_j$,
\begin{equation}
r_c(x_j)\approx r_*+\frac{x_j^2}{2}r_c^{\prime\prime}(0),
\end{equation}
where we have used the relation $r_c^\prime(0)=0$.

Also, the expansion for $a(r_j)$, keeping only terms up to second
order in $x_j^m\delta r_n$, reads
\begin{equation}
a(r_j)\approx a_*+\delta
ra^\prime_*+r_c^{\prime\prime}(0)a^\prime_*\frac{x_j^2}{2}+\frac{\delta
r^2}{2}a_*^{\prime\prime},
\end{equation}
where $a_*\equiv a(r_*)$, $a^\prime_*\equiv a^\prime(r_*)$, etc.
Thus, the exponent in (\ref{zs}) can be approximated as
\begin{equation}
S_j^{NG}=\frac{\mathcal{T}\Delta
x}{2\pi\alpha^\prime}\sqrt{C_1x_j^2+C_2},\label{zs1}
\end{equation}
with
\begin{equation}
C_1=\frac{r_c^{\prime\prime}(0)}{2}[2r_c^{\prime\prime}(0)+a_*^\prime],\qquad
C_2=a_*+\delta ra_*^\prime+\frac{\delta r^2}{2}a^{\prime\prime}_*.
\end{equation}

If the function in the square root of (\ref{zs1}) is negative,
$S_j^{NG}$ will contribute to an imaginary potential. The relevant
region of the fluctuations is the one between $\delta r$ that
yields a vanishing argument in the square root of (\ref{zs1}). So,
one can isolate the $j-$th contribution as
\begin{equation}
I_j\equiv\int_{\delta r_{jmin}}^{\delta r_{jmax}}d(\delta
r_j)exp[\frac{i\mathcal{T}\Delta
x}{2\pi\alpha^\prime}\sqrt{C_1x_j^2+C_2}],\label{ij}
\end{equation}
where $\delta r_{jmin}$ and $\delta r_{jmax}$ are the roots of
$C_1x_j^2+C_2$ in $\delta r$.

The integral in (\ref{ij}) can be calculated by using the saddle
point method for $\alpha^\prime<<1$ (the classical gravity
approximation). The exponent has a stationary point when the
function
\begin{equation}
D(\delta r_j)\equiv C_1x_j^2+C_2(\delta r_j),
\end{equation}
assumes an extremal value, and this happens for
\begin{equation}
\delta r=-\frac{a_*^\prime}{a_*^{\prime\prime}}.
\end{equation}

It is required that the square root has an imaginary part, results
in
\begin{equation}
D(\delta r_j)<0\rightarrow-x_c<x_j<x_c,
\end{equation}
with
\begin{equation}
x_c=\sqrt{\frac{1}{C_1}(\frac{a_*^{\prime2}}{2a_*^{\prime\prime}}-a_*)}.\label{xc}
\end{equation}

One takes $x_c=0$ if the square root in (\ref{xc}) is not real.
With these conditions, one can approximate $D\delta(r)$ by
$D(-\frac{a_*^\prime}{a_*^{\prime\prime}})$ in (\ref{ij})
\begin{equation}
I_j\sim exp[\frac{i\mathcal{T}\Delta
x}{2\pi\alpha^\prime}\sqrt{C_1x_j^2+a_*-\frac{a_*^{\prime2}}{2a_*^{\prime\prime}}}].\label{ij1}
\end{equation}

As the total contribution to the imaginary part comes from
$\Pi_jI_j$, one finds
\begin{equation}
ImV_{Q\bar{Q}}=-\frac{1}{2\pi\alpha^\prime}\int_{|x|<x_c}dx\sqrt{-x^2C_1-a_*+\frac{a_*^{\prime2}}{2a_*^{\prime\prime}}}.\label{im0}
\end{equation}

Integrating (\ref{im0}), one ends up with the imaginary potential
for the transverse case as
\begin{equation}
ImV_{Q\bar{Q}}=-\frac{1}{2\sqrt{2}\alpha^\prime}(\frac{a_*^\prime}{2a_*^{\prime\prime}}-\frac{a_*}{a_*^\prime}).
\label{im}
\end{equation}

Actually, there are several restrictions on Eq.(\ref{im})
\cite{JN2}. First, the imaginary potential should be negative, so
that
\begin{equation}
\frac{a_*^\prime}{2a_*^{\prime\prime}}-\frac{a_*}{a_*^\prime}>0,
\label{c2}
\end{equation}
with
\begin{equation}
a_*^\prime=e^{2P_*}(H_*^\prime +2H_*P_*^\prime), \label{a1}
\end{equation}
\begin{equation}
a_*^{\prime\prime}=e^{2P_*}(H_*^{\prime\prime}+4H_*^\prime
P_*^\prime+4H_*P_*^\prime
P_*^\prime+2H_*P_*^{\prime\prime}),\label{a2}
\end{equation}
which leads to
\begin{equation}
\varepsilon>\varepsilon_{min},
\end{equation}
with $\varepsilon=1/(4r_*-3)$.

The second restriction is related to the maximum value of
$DT_{max}$. To address this, we present $DT$ versus $\varepsilon$
for some choices of $b$ in the left panel of fig.4. One can see
that in each plot there exists a maximum value of $DT_{max}$. As
stated in \cite{JN2}, $DT_{max}$ indicates the limit of the saddle
point approximation: to go to higher $DT$, there are other string
configurations that may contribute to the calculation of the
Wilson loops besides the semiclassical U-shaped string
configuration \cite{DB}, implying that one cannot calculate the
imaginary potential in this case. On the other hand, we need to
take $\varepsilon<\varepsilon_{max}$, where $\varepsilon_{max}$ is
the maximum value of $\varepsilon$ for which the connected
contribution is valid. Thus, the domain of applicability of
Eq.(\ref{im}) is
$\varepsilon_{min}<\varepsilon<\varepsilon_{max}$.

Let us discuss results. First, we study the effect of the magnetic
field on the inter-distance. From the left panel of fig.4, one can
see clearly that increasing $b$ leads to decreasing $DT$. Namely,
the inclusion of a magnetic field decreases the inter-distance,
consistently with the findings of \cite{RRO}.

Next, we analyze the influence of a magnetic field on the
imaginary potential. After considering
$\varepsilon_{min}<\varepsilon<\varepsilon_{max}$, we compute the
imaginary potential from (\ref{im}), (\ref{a1}) and (\ref{a2}). In
the right panel of fig.4, we plot Im$V/(\sqrt{\lambda}T)$ against
$DT$ for $b=0,1,2,3$, corresponding to $\mathcal{B}/T^2\simeq
0,19,60,349$, respectively. From the figures one can see that for
each plot the imaginary potential starts at a $DT_{min}$,
corresponding to $\varepsilon_{min}$, and ends at a $DT_{max}$,
corresponding to $\varepsilon_{max}$. And increasing $b$, the
imaginary potential is generated for smaller distance. Namely, the
magnetic field enhances the imaginary potential. As we know, the
dissociation properties of quarkonia should be sensitive to the
imaginary potential, and if the onset of the imaginary potential
happens for smaller $DT$, the suppression will be stronger
\cite{JN2}. On the other hand, a larger imaginary potential
corresponds to smaller thermal width \cite{JN,JN1,KB}. Therefore,
one concludes that the presence of the magnetic field tends to
enhance the imaginary potential thus decreasing the thermal width.
One step further, the magnetic field may make the suppression
stronger. Intriguingly, it was argued \cite{RRO} that the magnetic
field has the effect of increasing the heavy quark potential thus
decreasing the dissociation length, in agreement with the findings
here.

\begin{figure}
\centering
\includegraphics[width=8cm]{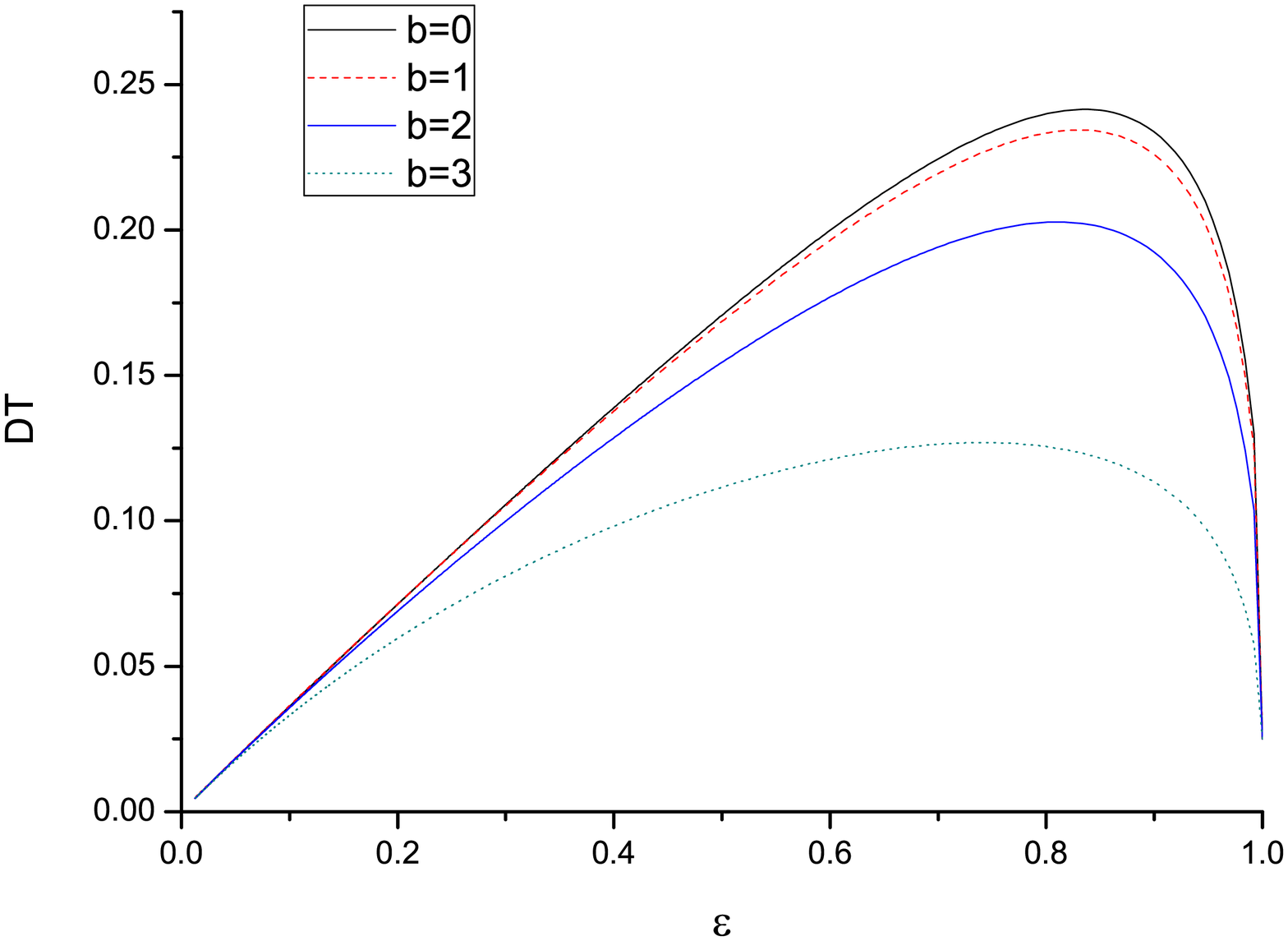}
\includegraphics[width=8cm]{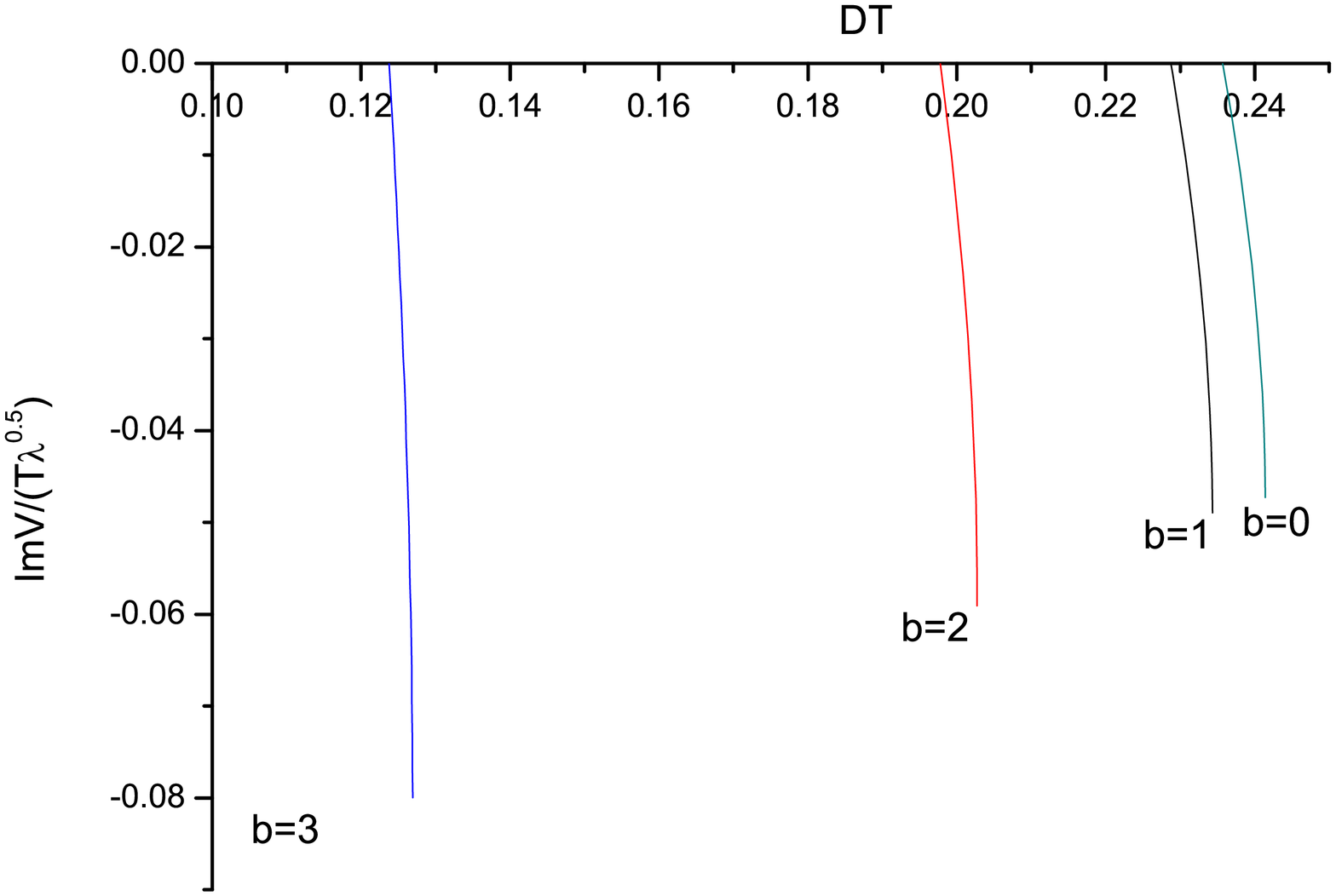}
\caption{ Left: $DT$ versus $\varepsilon$ for $\theta=\pi/2$. From
top to bottom $b=0,1,2,3$, respectively. Right:
$ImV/(\sqrt{\lambda}T)$ versus $DT$ for $\theta=\pi/2$. From right
to left $b=0,1,2,3$, respectively.}
\end{figure}

\subsection{Parallel to the magnetic field ($\theta=0$)}
Next, we discuss the $Q\bar{Q}$ axis parallel to the magnetic
field in the $z$ direction. The coordinates are parameterized as
\begin{equation}
t=\tau, \qquad x=0,\qquad y=0,\qquad z=\sigma,\qquad
r=r(\sigma),\label{par1}
\end{equation}
where the quark and anti-quark are located at $z=-\frac{D}{2}$ and
$z=\frac{D}{2}$, respectively.

The next analysis is similar to the previous subsection, so we
just show the final results. The inter-distance is
\begin{equation}
D=2\int_{r_*}^{\infty}dr\sqrt{\frac{A(r_*)}{A^2(r)-A(r)A(r_*)}}\label{x1}.
\end{equation}
with
\begin{equation}
A(r)=H(r)e^{2K(r)},\qquad A(r_*)=H(r_*)e^{2K(r_*)}.
\end{equation}

The imaginary potential is
\begin{equation}
ImV_{Q\bar{Q}}=-\frac{1}{2\sqrt{2}\alpha^\prime}[\frac{A_*^\prime}{2A_*^{\prime\prime}}-\frac{A_*}{A_*^\prime}],\label{im1}
\end{equation}
with
\begin{equation}
A_*^\prime=e^{2K_*}(H_*^\prime +2H_*K_*^\prime), \label{a11}
\end{equation}
\begin{equation}
A_*^{\prime\prime}=e^{2K_*}(H_*^{\prime\prime}+4H_*^\prime
K_*^\prime+4H_*K_*^\prime
K_*^\prime+2H_*K_*^{\prime\prime}).\label{a21}
\end{equation}

To study the effect of a magnetic field on the imaginary potential
for the parallel case, we plot $DT$ versus $\varepsilon$ and
Im$V/(\sqrt{\lambda}T)$ versus $DT$ in fig.5. From these figures,
one can see that the behavior is very similar to the transverse
case: the presence of the magnetic field decreases the
inter-distance and enhances the imaginary potential. In addition,
by comparing fig.4 and fig.5, one can see that the magnetic field
has a stronger effect for the transverse case rather than
parallel. Interestingly, a similar observation has been found in
\cite{RRO} which argues that the magnetic field has stronger
effect on the heavy quark potential for the perpendicular
configuration.

\begin{figure}
\centering
\includegraphics[width=8cm]{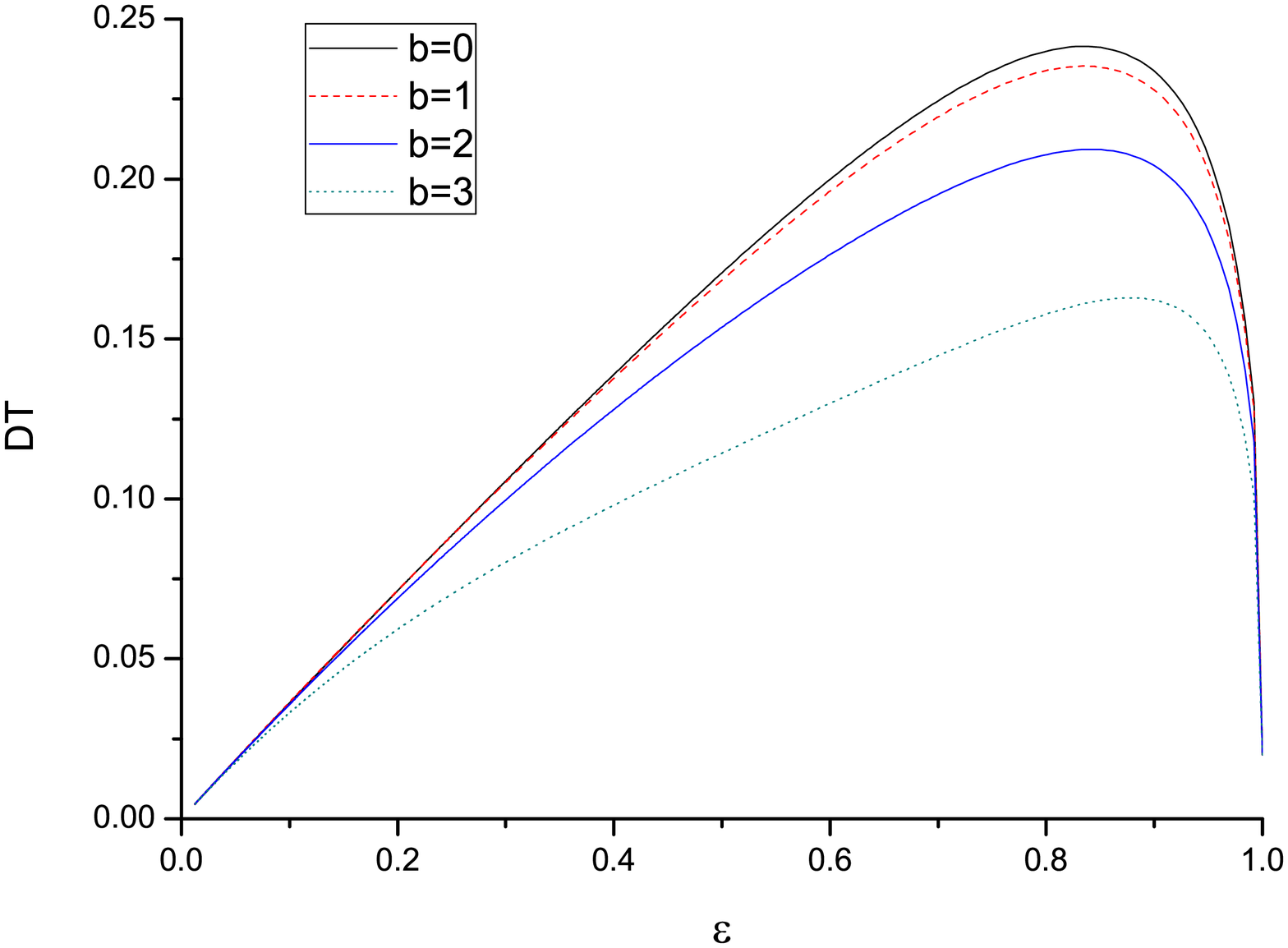}
\includegraphics[width=8cm]{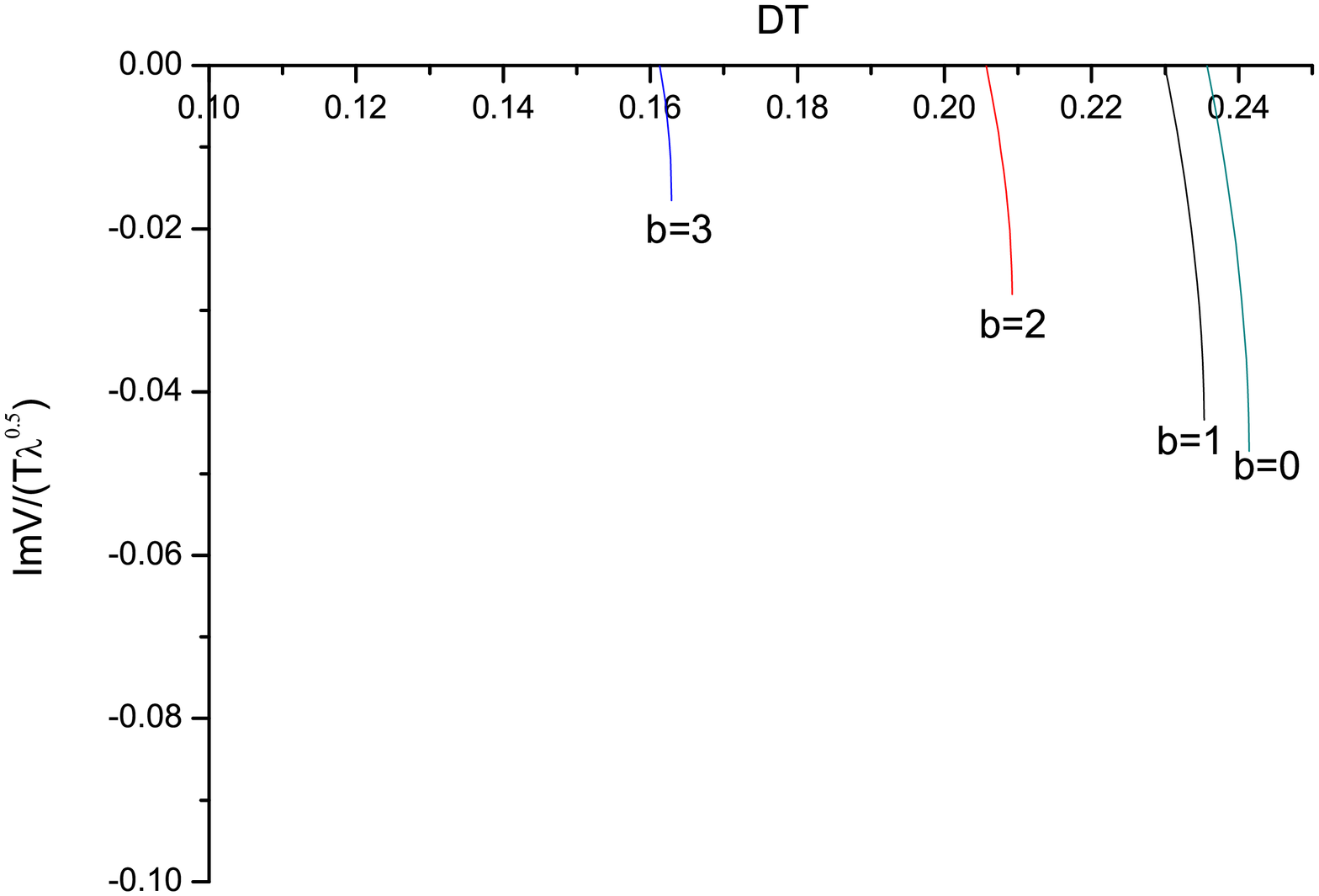}
\caption{ Left: $DT$ versus $\varepsilon$ for $\theta=0$. From top
to bottom $b=0,1,2,3$, respectively. Right:
$ImV/(\sqrt{\lambda}T)$ versus $DT$ for $\theta=0$. From right to
left $b=0,1,2,3$, respectively.}
\end{figure}

\section{conclusion and discussion}
In this paper, we studied the effect of a constant magnetic on the
imaginary potential of quarkonia in strongly-coupled
$\mathcal{N}=4$ SYM plasma. We considered the pair axis to be
aligned perpendicularly and parallel to the magnetic field, in
turn. In both cases, it is shown that the inclusion of the
magnetic field enhances the imaginary potential thus decreasing
the the thermal width. Also, the magnetic field has a stronger
effect for the perpendicular case rather than parallel.

However, it should be noted that the magnetized plasma considered
here is different from the real QGP, mainly due to the lack of a
dynamical breaking of conformal symmetry. It would be interesting
to investigate modifications of the holographic setup addressed
here and consider the systems that are not conformal. We leave
this for further study.

\section{Acknowledgments}
The authors would like to thank the anonymous referee for his/her
valuable comments and helpful advice. This work is partly
supported by the Ministry of Science and Technology of China
(MSTC) under the ¡°973¡± Project No. 2015CB856904(4). Z-q Zhang is
supported by NSFC under Grant No. 11705166. D-f. Hou is partly
supported by the NSFC under Grants Nos. 11735007, 11521064.



\begin{thebibliography}{0}

\bibitem{JA}
J. Adams et al. [STAR Collaboration], Nucl. Phys. A 757, 102
(2005).

\bibitem{KA}
K. Adcox et al. [PHENIX Collaboration], Nucl. Phys. A 757, 184
(2005).

\bibitem{EV}
E. V. Shuryak,  Nucl. Phys. A 750, 64 (2005).


\bibitem{TMA}
T. Matsui, H. Satz, Phys. Lett. B 178, 416 (1986).

\bibitem{ML}
M. Laine, O.Philipsen, P. Romatschke and M. Tassler, JHEP 03
(2007) 054

\bibitem{AB}
A. Beraudo, J.-P. Blaizot, and C. Ratti, Nucl. Phys. A806, 312
(2008).

\bibitem{NB}
N. Brambilla, J. Ghiglieri, A. Vairo, and P. Petreczky, Phys. Rev.
D 78, 014017 (2008).

\bibitem{MAE}
M. A. Escobedo, J. Phys. Conf. Ser. 503 (2014) 012026.

\bibitem{NB1}
N. Brambilla, M.A. Escobedo, J. Ghiglieri, J. Soto and A. Vairo,
JHEP 09 (2007) 038


\bibitem{AD}
A. Dumitru, Y. Guo, and M. Strickland, Phys.Rev. D79 (2009)
114003.

\bibitem{MM}
M. Margotta, K. McCarty, C. McGahan, M. Strickland, and D. Y.
Elorriaga,  Phys.Rev. D 83 (2011) 105019.

\bibitem{VC}
V. Chandra and V. Ravishankar,  Nucl. Phys. A 848 (2010) 330.


\bibitem{Maldacena:1997re}
J. M. Maldacena,  Adv. Theor. Math. Phys. 2, 231 (1998).

\bibitem{Gubser:1998bc}
S. S. Gubser, I. R. Klebanov and A. M. Polyakov,  Phys. Lett.
B428, 105 (1998).

\bibitem{MadalcenaReview}
O. Aharony, S. S. Gubser, J. Maldacena, H. Ooguri and Y. Oz, Phys.
Rept. 323, 183 (2000).

\bibitem{JCA}
J. C. Solana, H. Liu, D. Mateos, K. Rajagopal, and U. A.
Wiedemann, arXiv:1101.0618.

\bibitem{JN}
J. Noronha, A. Dumitru,  Phys. Rev. Lett. 103 (2009) 152304.

\bibitem{JN1}
S. I. Finazzo, J. Noronha, JHEP 11 (2013) 042.

\bibitem{KB}
K. B. Fadafan, D. Giataganas and H. Soltanpanahi, JHEP 11 (2013)
107

\bibitem{JN2}
S. I. Finazzo, J. Noronha, JHEP 01 (2015) 051.



\bibitem{MAL}
M. Ali-Akbari, D. Giataganas and Z. Rezaei, Phys. Rev. D 90 (2014)
086001.

\bibitem{KB1}
K. B. Fadafan, S. K. Tabatabaei, J. Phys. G: Nucl. Part. Phys. 43
095001 (2016).


\bibitem{ZQ}
Z. q. Zhang, D.f. Hou, and G. Chen, Phys. Lett. B 768, 180 (2017).


\bibitem{NR}
N. F. Braga and L. F. Ferreira, Phys. Rev. D 94, 094019 (2016).


\bibitem{JS}
J. Sadeghi and S. Tahery, JHEP 06 (2015) 204.



\bibitem{JLA}
J. L. Albacete, Y. V. Kovchegov, A. Taliotis, Phys. Rev. D 78
(2008) 115007.

\bibitem{THA}
T. Hayata, K. Nawa, T. Hatsuda, Phys. Rev. D 87 (2013) 101901.


\bibitem{ED}
E. D. Hoker and P. Kraus, JHEP 10 (2009) 088.


\bibitem{KAMA}
K. A. Mamo,  JHEP 08 (2013) 083.


\bibitem{RCR}
R. Critelli, S. I. Finazzo, M. Zaniboni, and J. Noronha, Phys.
Rev. D 90, 066006 (2014).


\bibitem{RRO}
R. Rougemont, R. Critelli and J. Noronha, Phys. Rev. D 91, 066001
(2015).

\bibitem{KIM}
K. A. Mamo, Phys. Rev. D 94, 041901(R) (2016).



\bibitem{SLI}
S. Li, K. A. Mamo and H.U.Yee, Phys. Rev. D 94, 085016 (2016).


\bibitem{JMM}
J. M. Maldacena, Phys. Rev. Lett. 80, 4859 (1998).

\bibitem{ABR}
A. Brandhuber, N. Itzhaki, J. Sonnenschein and S. Yankielowicz,
Phys. Lett. B 434, 36 (1998).

\bibitem{SJR}
S. -J. Rey, S. Theisen and J. -T. Yee, Nucl. Phys. B 527, 171
(1998).

\bibitem{DB}
D. Bak, A. Karch, L. G. Yaffe, JHEP 0708 (2007) 049.





\end{thebibliography}
\end{document}